\documentclass[twocolumn]{aastex62}

\usepackage{apjfonts}
\usepackage[T1]{fontenc}
\usepackage{color}
\usepackage{amsmath,amstext}
\usepackage{hyperref}
\usepackage{natbib}
\usepackage{graphicx}
\usepackage{gensymb}
\usepackage{float}

\begin{document}

\title{\sc Single-pulse gamma-ray bursts have prevalent hard-to-soft spectral evolution} 

\author{Ian Busby}
\affil{ Department of Physics, Oregon State University, 301
  Weniger Hall, Corvallis, OR 97331, USA}

\author{Davide Lazzati}
\affil{ Department of Physics, Oregon State University, 301
  Weniger Hall, Corvallis, OR 97331, USA}

%%%%%%%%%%%%%%%%%%%%%%%%%%%%%%%%%%%%%%%%%%%%%%%%%%
\begin{abstract}
We analyze the spectral evolution of 62 bright Fermi gamma-ray bursts with large enough signal to noise to allow for time resolved spectral analysis. We develop a new algorithm to test for single-pulse morphology that is insensitive to the specific shape of pulses. Instead, it only checks whether or not there are multiple, isolated, statistical significant peaks in the light curve. In addition, we carry out a citizen science test to assess light curve morphology and spectral evolution.
We find that, no matter the adopted assessment method, bursts characterized by 
single-peaked prompt emission light curves have a greater tendency to also have a consistently decaying peak energy, or hard-to-soft spectral evolution. This contrasts the behavior of multi-peaked bursts, for which the tendency is to have a peak frequency that is not monotonically decreasing. We discuss this finding in the theoretical framework of internal/external shocks, and find it to be consistent with at least some single pulse bursts being associated with particularly high-density environments.
\end{abstract}

%%%%%%%%%%%%%%%%%%%%%%%%%%%%%%%%%%%%%%%%%%%%%%%%%%


\section{Introduction}
\label{intro}

Gamma-Ray Bursts (GRBs), the brightest explosions in the present universe \citep{Kulkarni1999}, have been the subject of intense study for more than 50 years \citep{Klebesadel1973}. Many discoveries and theoretical breakthroughs have allowed for the establishment of a standard model to interpret the variety of observations. In this model, all bursts are cosmological in origin \citep{Fishman1995} and there are two predominant classes: short bursts that last less than about 2 s and long bursts that last more than approximately 2 seconds \citep{Kouveliotou1993}. All bursts are characterized by the presence of a central engine, either a black hole or a neutron star, that releases a relativistic, possibly magnetized outflow. Long bursts are associated  with the core collapse of massive, compact, and fastly rotating stars, their duration set by the accretion time of bound stellar material on an accretion disk surrounding the engine \citep{Galama1998,Popham1999,Lee_2000a,Kohri_2002,Hjorth2003,Woosley2006,Chen_2007,2017NewAR..79....1L}. Short bursts, instead, are associated with the merger of compact binary systems either made of two neutron stars or, perhaps, by a neutron star and a black hole \citep{Eichler1989,Abbott2017a,Abbott2017b,Lazzati2018,Mooley2018,Ghirlanda2019}. In their case, the burst duration is expected to be driven by the viscous timescale of the accreting material. While not all bursts clearly fit into this scenario (e.g., \citealt{Rastinejad2022}), it is a model that has had significant success in accounting for most observed properties of both individual bursts and of the ensemble of several thousand observed events \citep{Gehrels2009rev}.

One property of bursts that has so far eluded a robust interpretation and classification is their spectral behavior \citep{Band1993,Preece2000,Gruber2014}. Most bursts are characterized by a non-thermal, broad band spectrum \citep{Band1993}. The hardness ratio of short bursts is higher than that of long ones \citep{Kouveliotou1993}. Additional information can be extracted by looking at the evolution of the peak frequency of the spectrum $\epsilon_{\rm{peak}}$. This is defined as the photon energy where the $\nu F(\nu)$ function peaks. Also in this case the behavior of $\epsilon_{\rm{peak}}$ during the prompt phase of the burst defines two classes. In most events $\epsilon_{\rm{peak}}$ tracks the luminosity (e.g., \citealt{Golenetskii1983}): it grows at the beginning of a pulse, peaks when the luminosity is the highest, and decreases afterwards. If a burst is characterized by multiple pulses, this tracking repeats for as many pulses as are observed. This is also true across different bursts, as testified by the existence of the Golenetskii correlation \citep{Golenetskii1983,Lu2012}. A second class of events, instead, displays a consistently monotonic hard-to-soft behavior, in which the peak energy of the photons decreases with time, irrespective of the burst luminosity and of whether the burst is characterized by a single or multiple pulses (e.g., \citealt{Norris1986}). The origin of these two classes is unclear, partly because it can be studied only for the subclass of bright bursts, for which a time-resolved spectral analysis can be carried out.

Recently, \cite{Lazzati:2022mgg} have proposed a possible origin for bursts with hard-to-soft evolution. In their model some bursts take place within the accretion disk of supermassive black holes at the center of their host galaxies, and are therefore embedded in a gas that is many orders of magnitude denser than the interstellar medium. \cite{Lazzati:2022mgg} show that in that case the burst prompt emission is not produced by  either a photospheric or internal shock component but rather  by the early onset of the external shock. The pulses of these bursts would inevitably be longer than their separation, merging in a single, possibly undulating observed pulse. Because pulse peak frequency would decrease with time (like in the afterglow), the envelope pulse would display hard-to-soft evolution, even during rebrightenings and in the initial growing phase of the pulse. A clear prediction of their model is therefore that hard-to-soft evolution should be preponderant in single pulse bursts, while the tracking behavior should instead predominantly be observed in multi-pulse bursts.

In this paper we study a sample of bright bursts for which time-resolved spectroscopy was carried out looking for evidence of such an imbalance of spectral evolution types. This paper is organized as follows: in Section~2 we present the sample of burst and the techniques used to classify bursts both in the temporal and spectroscopic domains. In Section~3 we present our results and discuss their statistical significance. Finally, in Section~4 we discuss our results and possible strategies to improve their significance.

%%%%%%%%%%%%%%%%%%%%%%%%%%%%%%%%%%%%%%%%%%%%%%%%%%
\section{Methods}
\label{sec:methods}

\subsection{The sample}
The sample of bursts analyzed in this work was collected from the time-resolved spectra catalogue produced by \cite{Yu:2016epf}. Of the 81 bursts in the catalogue, 62  were utilized in this study. \cite{Yu:2016epf} used CSPEC data for 15 of the bursts, which they note has lower temporal resolution with higher spectral resolution, and a duration of around 8000 seconds. The other bursts in the catalogue were created using TTE data. For the creation of light-curves in this study TTE data was used. TTE has a duration of around 300 seconds \citep{Meegan_2009}. \cite{Yu:2016epf} fit each burst on multiple time intervals. Four different spectral models were used: BAND, COMP, SBPL, and power law \citep{Yu:2016epf}. The Band function (BAND; \citealt{Band1993}) is a four parameter piece-wise function with an exponentially smooth transition between two power laws \citep{Yu:2016epf}. The smoothly broken power law (SBPL; \citealt{Ryde1999,Kaneko2006}) is also a smooth transition between two asymptotic power-law behaviours. In general, SBPL has a parameter that controls the length and smoothness of the transition. In the  \cite{Yu:2016epf} catalogue, this parameter is fixed at 0.3. The Comptonized model (COMP) is a 3 parameter power law fit with an exponential cutoff. Finally, the power law fit is a simple 2 parameter power law. Additional details about the fits used can be found in \citep{Yu:2016epf} and the references therein. For a given interval of a given burst, \cite{Yu:2016epf} provides the best functional fit in their catalogue. Thus a burst may have different functional fits for different time intervals. Peak energy was of principle interest for this work, thus any spectral fit that did not include a peak energy value was omitted. In particular, the power law fits have no defined peak energy, and therefore any fit using the power law could not be used.  All bursts with at least 4 spectral fits that included well-defined peak energy values were included. Of the 19 excluded, 16 were excluded due to having too few spectra with peak energy values. Due to the difference in duration for data used, some GRBs were excluded as a majority of spectra occurred after or before the TTE data. The identification numbers of excluded bursts and the reason for their exclusion are reported in Table \ref{exclusion table}.

\begin{figure*}[!ht]
\begin{center}
\includegraphics[width=\textwidth]{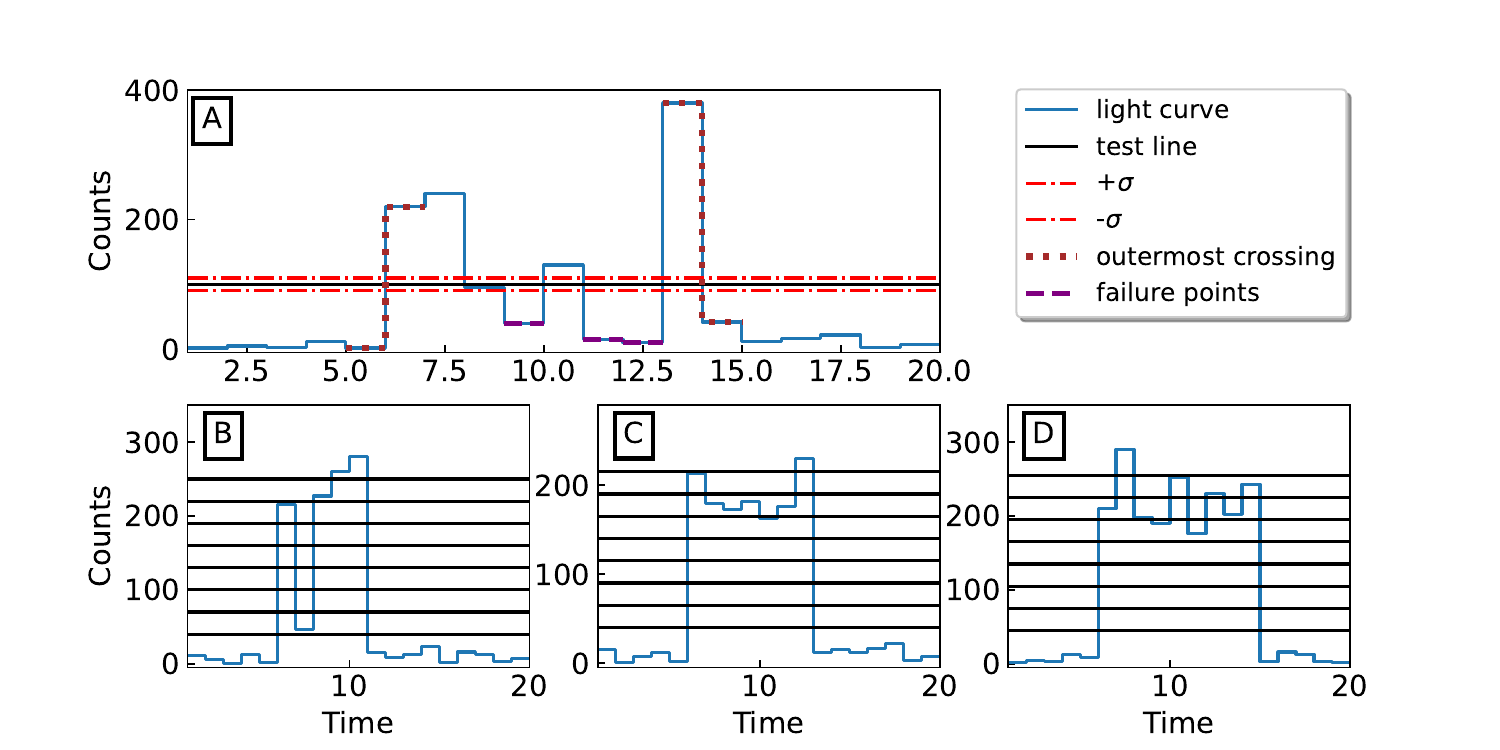}
\caption{A$)$ Pseudo light-curve to show an example of a single horizontal line test. Between the two outermost crossings, 3 failure points are found. B,C,D$)$ Three different pseudo light-curves with a horizontal line test consisting of 8 lines. B$)$ represents a light-curve with a single deep trough. C$)$ represents a single shallow, wide trough. D$)$ represents multiple small troughs. All three light-curves have 5 points in their total score.}
\label{fig:hozlineexample}
\end{center}
\end{figure*}

While we used the spectral fits from \cite{Yu:2016epf}, we re-accumulated burst light curves to ensure consistency and a format suitable for our horizontal line technique (see below). Light-curve TTE data was collected from the HEASARC BROWSE GBM Burst Catalogue \citep{Gruber_2014,Bhat_2016,vonKienlin_2014,vonKienlin:2020xvz} \footnote{\tt https://heasarc.gsfc.nasa.gov/W3Browse/fermi/fermigbrst.html}.
From the 12 NaI detectors the two with the brightest signals were chosen. For each burst we define a \emph{region of interest} as:
\begin{equation}\label{eqn:regionofinterest}
    \max({T_{90start} - 0.5T_{90},T_{dstart}}) \le t \le \min({T_{90start} +2T_{90},T_{end}})
\end{equation}
where $T_{90start}$ is the start of the $T_{90}$ period, $T_{dstart}$ is the earliest available TTE data, and $T_{end}$ is the end of the TTE data. If the burst is sufficiently short, the region of interest lasts $2.5T_{90}$. Unfortunately, not all bursts have TTE data coverage over the entire $2.5T_{90}$ period. TTE includes data from 30 seconds prior to the trigger time and lasts for around 300 seconds \citep{Meegan_2009}. In the case that the $T_{90start}-0.5T_{90}$ occurred before 30 seconds prior to trigger time, the earliest TTE data was used as the beginning of the region of interest. Similarly, if $T_{90start} + 2T_{90}$ occurred after the 300 second duration, the last TTE time was used as the end of the region of interest. 
For all bursts, the region of interest was divided into 120 equal length bins. Broadly, the length of each bin was the length of the region of interest for the burst divided by 120. For bursts where the $2.5T_{90}$ fell within the TTE data, bin duration was given simply by ${T_{bin}} = \frac{T_{90}}{48}$. This method ensured that the primary burst behavior was well described in the light curve.

\begin{table}[!t]
\centering
\label{exclusion table}
\begin{tabular}{|l|l|}
\hline
GRB       & Exclusion Reason                          \\ \hline
090804940  & Not Enough Spectra                        \\ \hline
100122616 & Not Enough Spectra                        \\ \hline
100511035 & Not Enough Spectra                        \\ \hline
100612726 & Not Enough Spectra                        \\ \hline
100722096 & Not Enough Spectra                        \\ \hline
100829876 & Not Enough Spectra                        \\ \hline
100910818 & Not Enough Spectra                        \\ \hline
101231067 & Not Enough Spectra                        \\ \hline
110428388 & TTE photon data missing                   \\ \hline
110729142 & Spectra outside TTE range 
\\ \hline
110817191 & Not Enough Spectra                        \\ \hline
110903009 & Not Enough Spectra                        \\ \hline
120624933 & Spectral data prior to TTE data            \\ \hline
081009140  & Not Enough Spectra                        \\ \hline
081124060  & Not Enough Spectra                        \\ \hline
081221681  & Not Enough Spectra                     
\\ \hline
110213220 & Not Enough Spectra                        \\ \hline
111127810 & Not Enough Spectra                        \\ \hline
111228657 & Not Enough Spectra                        \\ \hline
\end{tabular}
\caption{Excluded GRBs with reasons stated. GBM number is given in the left column.}
\end{table}

\subsection{Analysis}
In order to classify the light-curves and time resolved spectra we developed two independent techniques. The first is an exercised in citizen science, in which individuals from the public were asked to classify light curves and spectra. The second, instead, is a set of computer algorithms that automatically classify burst behavior.

In the citizen science project 24 participants were asked to rank all 62 light-curves and spectra. Light-curves and spectra were unlabelled and given to participants separately to minimize bias. Participants classified on a ternary system. Each light-curve was classified as a FRED, unsure, or not a FRED\footnote{Note that we used the terminology FRED (Fast Rise, Exponential Decay) even though individuals were not asked to evaluate the specific decay shape of the pulses}. Similarly the spectra were classified as "hard-to-soft", unsure, or not "hard-to-soft". Prior to classification participants were given an instruction sheet that gave brief descriptions of the meanings of FRED and hard-to-soft, as well as example light-curves from outside the data set. For both sets of data, a score of "yes" was assigned a value of 1, "unsure" a value of 0.5, and "no" a value of 0. The final score for a burst was the average score across all 24 participants. The uncertainty on the scores were taken using the standard error of the mean. 

Computational methods were also used to determine if given light-curves and time resolved spectra were single-peaked and hard-to-soft, respectively. 
By definition, a set of time resolved spectra being hard-to-soft means that the peak frequency for a given time is not larger than the peak frequency for all previous time steps.  Let $\epsilon_{peak,i}$ be the peak energy at some time index $i$ and let $\sigma_{\epsilon,i}$ be the error in peak energy at time index $i$. Let $j$ be some time step before $i$. To determine if a given peak frequency is not larger than the previous points, the difference in peak energy should be negative within one uncertainty. In other words, 
\begin{equation}\label{hts eq.1}
\epsilon_{peak,i}-\epsilon_{peak,j} - \sqrt{\sigma_{\epsilon_i}^2+\sigma_{\epsilon_j}^2}<0, \quad j<i 
\end{equation}
should be true for all peak energies of index $j$, where $j<i$. Code was produced that systematically verified that Equation \ref{hts eq.1} was true for every time value $i$, for all $j$ before $i$. While testing a given burst, each time a spectra of the burst fails this test a point is added to a running total for that burst. After running the test on all peak energy values for that burst, the hard-to-soft score is the total number of times the burst violates Equation \ref{hts eq.1}. This score, however, grows with the number of peak energy values. The total number of tests ran for $N$ peak energy values is $1+2+...+(N-1)$, or $\frac{N(N-1)}{2}$. To normalize for the availability of peak energies, the score for a given burst is thus divided by $\frac{N(N-1)}{2}$. Finally, to place the test on a more meaningful scale, the normalized score was subtracted from one, so that a hard-to-soft burst would correspond to a score close to one.

To test for single-peak behavior we developed a test in which a series of horizontal lines was used. To understand the logic of the test, consider a burst with a single peak. If a horizontal line is placed along the burst, all the points of the burst which are above the line should be in sequence. In other words, a single peak should only have one upward crossing and one downward crossing. Taking advantage of this fact, a set of sixteen equally spaced horizontal lines was placed on each burst. The highest line was placed two count-rate standard errors below the peak count rate. Photon counts follow a Poisson distribution thus the standard error for a count $N$ is $\sqrt{N}$. The average background count  was determined by taking the median of all counts outside the region defined by $T_{90start}<t<T_{90start}+T_{90}$.  The lowest line was placed two uncertainty above the background. This lowest line was excluded from the test, to ensure that any unusual background behavior would not affect the test. In total,  fifteen lines affected the score in our test. For a given line, we determined the set of points which were at least one uncertainty above the test line. The test then determines if between the first and last crossing any points fell at least 1 uncertainty below the line. See Figure \ref{fig:hozlineexample} for an example line and a few light curve cases that yield similar scores. For every point that fell at least 1 uncertainty below the line, a tally was added to a running total for that burst. This test was run on all fifteen lines. To normalize the scores, they were divided by 1800. This comes from the fact that there were 15 lines and 120 bins on the region of interest. If there were perfect delta spikes at both ends of the region of interest, the score would then be $15\cdot120$, giving 1800. For a more general version, a test consisting of $m$ lines with $N_{bins}$ bins on the region of interest would be normalized by $m\cdot N_{bins}$.  Finally, the normalized score was subtracted from one. This places the scores on a scale from zero to one, where a score of one means that the burst was perfectly single peaked, no points violated the horizontal line test. Conversely, a score of 0 represents a burst with multiple extremely well separate narrow peaks. The virtue of using multiple lines is best seen in Figure \ref{fig:hozlineexample}. A single line may miss some behavior, such as in \ref{fig:hozlineexample}C where only one line catches the shallow valley behavior. This is similarly helpful for narrow, deep valleys, in which a single line would only show one point as violating the test, but the deep peak is caught by multiple lines, increasing the score as seen in \ref{fig:hozlineexample}B. This test is versatile in its ability to catch multiple different multi-peaked behaviors through the multiple lines. It is also powerful in not requiring the pulse to have any specific analytical description. 

%%%%%%%%%%%%%%%%%%%%%%%%%%%%%%%%%%%%%%%%%%%%%%%%%%

%%%%%%%%%%%%%%%%%%%%%%%%%%%%%%%%%%%%%%%%%%%%%%%%%%
\section{Results}
\label{sec:results}

\subsection{Comparing methods}
\begin{figure}[ht]
\begin{center}
\includegraphics[width=\columnwidth]{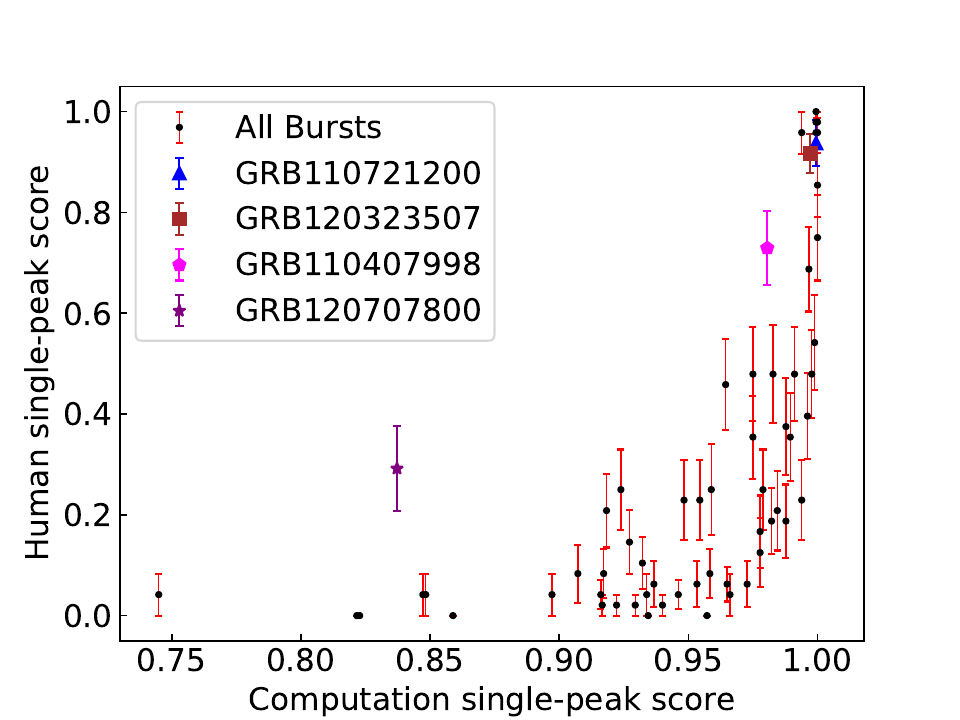}
\caption{ Human FRED score vs computational single peak score from the horizontal line test. Computational test was normalized by dividing by the number of lines times the number of bins on the region of interest. GRB 111220486 had a significantly lower computational score than all other bursts ([0.294,0.0833]) and lies outside the shown range. Notable outliers on both tests between the human and computational scores can be seen in the legend. }
\label{fig:fredhumvscomp}
\end{center}
\end{figure}

\begin{figure}[ht]
\begin{center}
\includegraphics[width=\columnwidth]{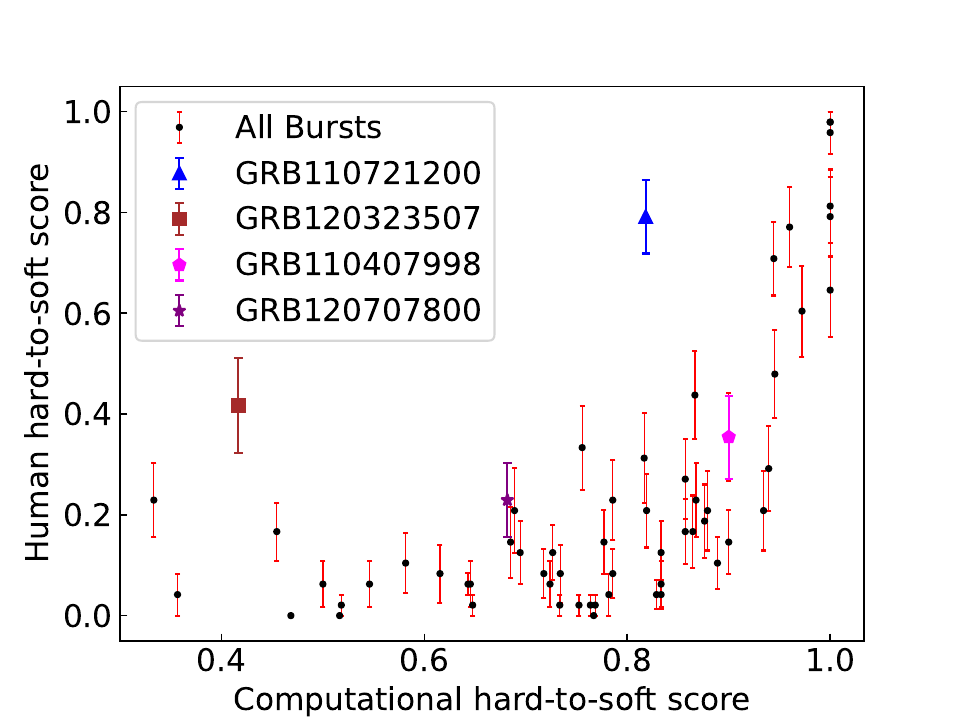}
\caption{ Human hard-to-soft score vs number of failure points on the hard-to-soft test. The hard-to-soft test goes over each point and compares it to all previous. As such, hard-to-soft failure points are normalized by dividing by $\frac{N(N-1)}{2}$ where N is the number of spectra. Notable outliers on both tests can be seen in the Legend.}
\label{fig:htshumvscomp}
\end{center}
\end{figure}

The human and computer rankings generally agreed with each other. Comparisons of the human classifications and computer classifications for both hard-to-soft and single peaked metrics can be seen in Figures \ref{fig:fredhumvscomp} and \ref{fig:htshumvscomp}, respectively. In both Figures \ref{fig:fredhumvscomp} and \ref{fig:htshumvscomp} there is a positive trend suggesting the tests agree. It is important to note that while both the human and computational tests are on a 0 to 1 scale, they do not hold the same meaning. For instance, it is common for the human single-peak test scores to be near 0, as this simply means a majority of the participants stated "no" for the classification. However, on the computational single-peaked test, a score of 0 would mean that the burst has well-separated delta-function like peaks. This is an ideal multi-peaked burst, and as such most bursts do not have scores close to 0 even with multiple-peaks. Similarly, scores for the computational hard-to-soft test would have to be strictly monotonically increasing which again is unlikely. Thus a score of 0 is again quite unlikely. This means while a score of 0.8 for the human single-peak test indicates the burst is single-peaked, a similar score on the computational single-peak test does not. This holds similarly true for the hard-to-soft test. For both the computational single-peak and hard-to-soft tests a score must be much closer to 1 for the burst to be considered single-peaked or hard-to-soft than in the human tests.

\begin{figure}[ht]
\begin{center}
\includegraphics[width=\columnwidth]{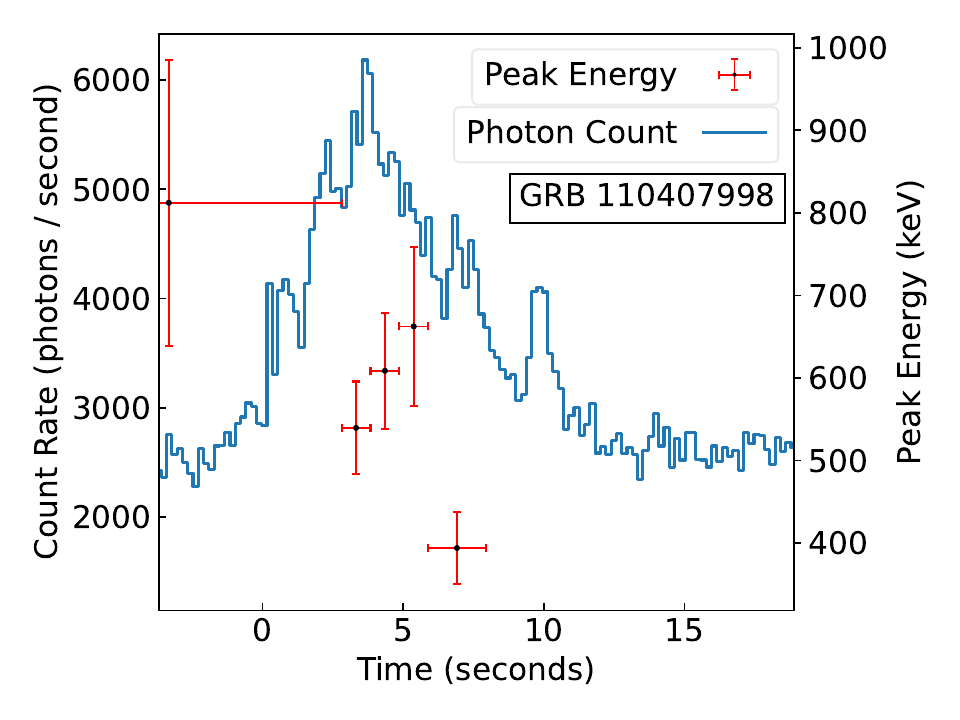}
\caption{GRB110407998, outlier in single peak horizontal line test. Light curve and peak energy values are shown. }
\label{fig:fredout}
\end{center}
\end{figure}

In order to quantitatively compare the results, both Pearson and Spearman tests were run to test for linear correlation and monotonic correlation, respectively. Table \ref{tab:hum vs comp stat scores} shows the Pearson and Spearman scores for both the single peak and hard-to-soft tests as well as the probability of a set of random data showing a correlation as strong or stronger. 

\begin{table}[!t]
\centering
\begin{tabular}{|l|l|l|l|l|}
\hline
                     & $r_p$   & $r_s$   & $p_{Pearson}$         & $p_{Spearman}$         \\ \hline
Single-peak & 0.3807 & 0.8088 & $2.267 \cdot 10^{-3}$ & $1.847 \cdot 10^{-15}$ \\ \hline
Hard-to-soft  & 0.5867 & 0.6502 & $5.413 \cdot 10^{-7}$ & $1.072 \cdot 10^{-8}$  \\ \hline
\end{tabular}
\caption{Computational vs Human Pearson and Spearman scores. Values labelled $r_p$ and $r_s$ represent the Pearson and Spearman correlation coefficients. Values labelled $p_{Pearson}$ and $p_{Spearman}$ indicate p-values for the Pearson and Spearman tests respectfully.}
\label{tab:hum vs comp stat scores}
\end{table}

\begin{figure}[ht]
\begin{center}
\includegraphics[width=\columnwidth]{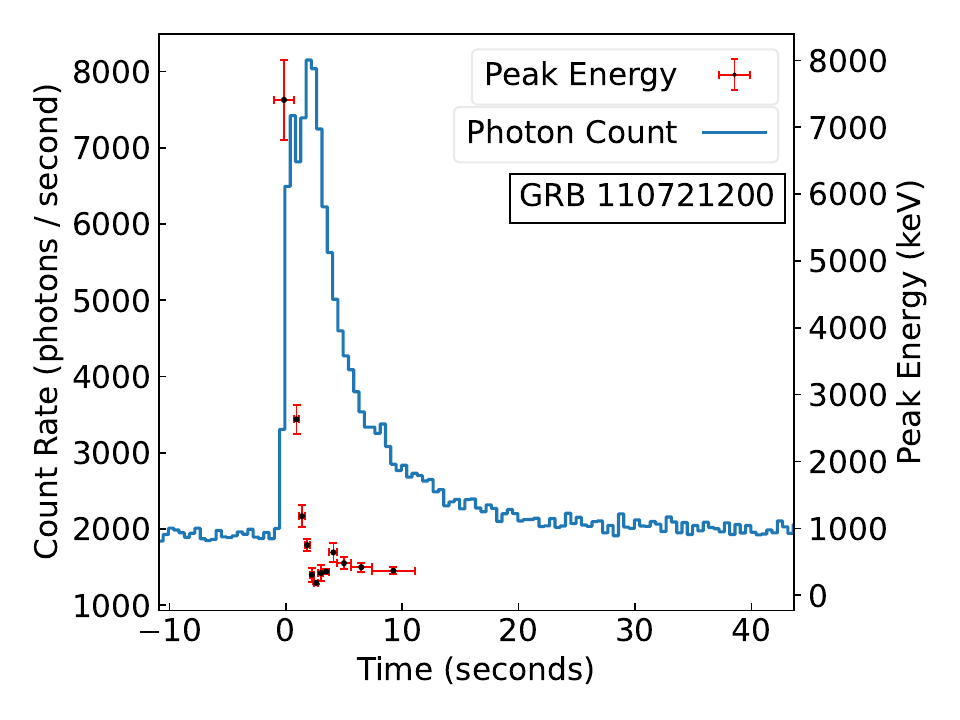}
\caption{GRB110721200, outlier in the hard-to-soft test. Light curve and peak energy values are shown.}
\label{fig:htsout}
\end{center}
\end{figure}

For both sets of tests, there was both a fairly strong Spearman and Pearson correlation with a high probability. This shows that the tests generally agreed with one another; a human scoring of hard-to-soft typically corresponded to a computational scoring of hard-to-soft and similarly for the single-peak test.

There are several outliers that can be seen in Figures \ref{fig:fredhumvscomp} and \ref{fig:htshumvscomp}. The most noticeable outlier for the single peak test is GRB110407998, whose light-curve and spectra are shown in Figure \ref{fig:fredout}. GRB110507998 was ranked as more single peaked than other bursts with similar computational scores. The light-curve seems to show a generally single peaked behavior aside from a small secondary peak at 10 seconds. The horizontal line test picked up on this second peak, whereas the human eye seemed to generally classify this as background. The second peak rises up nearly 1000 photons/s above the background, well above the uncertainty. 

The most noticeable outlier in the hard-to-soft tests is GRB110721200 whose light-curve and spectra can be seen in Figure \ref{fig:htsout}. GRB110721200 was generally considered much more hard-to-soft than other bursts with similar computational rankings. The spectra of GRB110721200 decrease until around 3 seconds before increasing slightly and then decreasing further. The computational hard-to-soft test picked up several failure points in the section of increasing peak energy around 3 seconds. Humans ranked this as hard-to-soft.

Except for these outliers, the computational and human results generally agreed. It is difficult to directly compare the results given the different scales for the tests. For instance, only a handful of bursts scored a single-peaked human score greater than 0.8, whereas almost all bursts had computational single peak scores above 0.8. That being said, from Figure \ref{fig:fredhumvscomp}, the human test seemed to have more bursts sitting in the range around 0.5, or the "unsure" range when the computational test would consider them multi-peaked. This may indicate that the computational test is stricter in its classification of bursts as single-peaked.  

\subsection{Correlation between light-curve and spectral evolution}

\begin{figure}[ht]
\begin{center}
\includegraphics[width=\columnwidth]{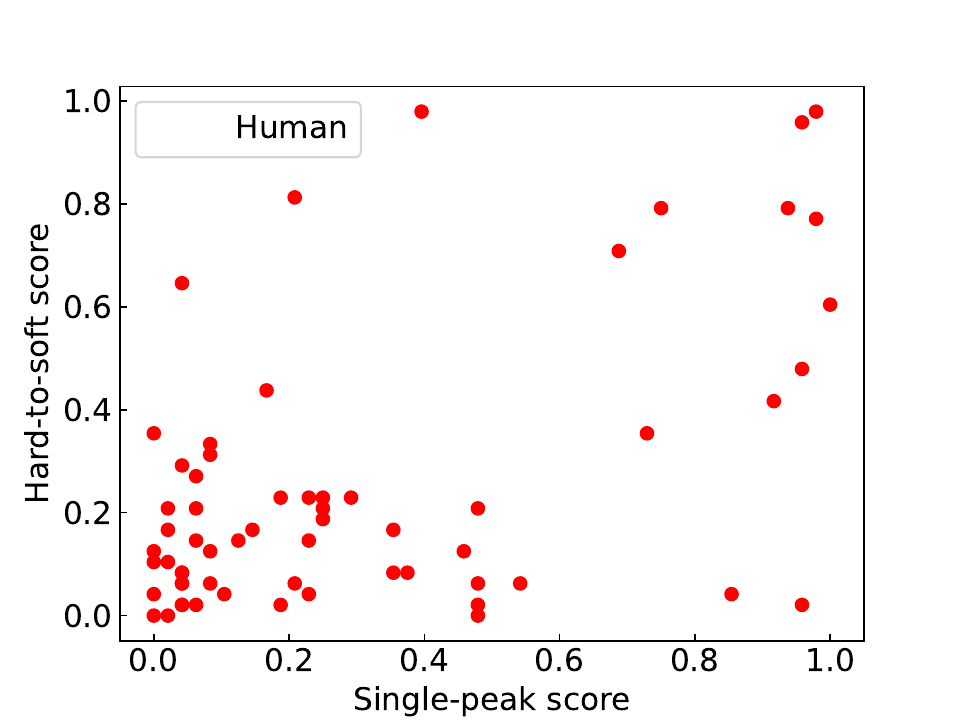}
\caption{Human results comparing hard-to-soft to single peak scores. Scoring was done on a yes (1) unsure (0.5) no (0) scale for both axes. Points represent the average score by all participants, and the standard error has been taken represented by the error bars.}
\label{fig:fhtshumanunbinned}
\end{center}
\end{figure}

\begin{figure}[ht]
\begin{center}
\includegraphics[width=\columnwidth]{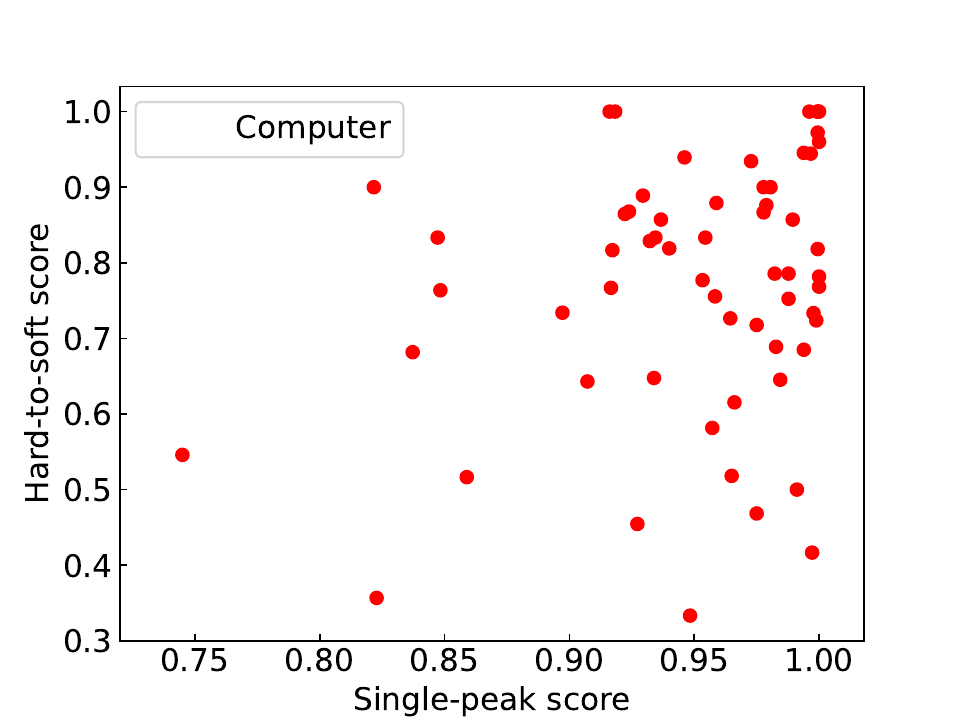}
\caption{Comparison of computational scores for the hard-to-soft and single peak tests. Both have been normalized. A larger score indicates the burst is less hard-to-soft or single peaked. GRB 111220486 had a significantly lower single-peak score ([0.219,0.694]) than all other bursts and lies outside the shown range.}
\label{fig:fhtscompunbinned}
\end{center}
\end{figure}

The hard-to-soft scores and single peaked scores were directly compared for both the computational and human methods. We will first consider the unbinned results. Figures \ref{fig:fhtscompunbinned} and \ref{fig:fhtshumanunbinned} show the human and computational results respectively. To test for potential correlations the Pearson and Spearman tests were again run. Scores and probability p-values can be seen in Table \ref{tab:hts vs fred tab results}. These tests are readily applicable, as they do not require uncertainty, and for neither the human nor computational tests are there well-defined uncertainties. 

\begin{table}[]
\centering
\begin{tabular}{|l|l|l|l|l|l|}
\hline
      & $r_{Pearson}$ & $p_{Pearson}$        & $r_{Spearman}$ & $p_{Spearman}$ & Sig diff \\ \hline
Comp & 0.1792        & 0.1634               & 0.2442         & 0.05580        & 3.326    \\ \hline
Human  & 0.5411        & $5.616\cdot 10^{-6}$ & 0.3600         & 0.004056       & 4.303    \\ \hline
\end{tabular}
\caption{Pearson and Spearman tests on the hard-to-soft vs single peak scores for both the computational and human tests. Last column describes number of uncertainty in the hard-to-soft scores for single-peaked vs multi-peaked bursts. A positive value indicates that single-peaked bursts hard-to-soft score was larger.}
\label{tab:hts vs fred tab results}
\end{table}

On both the Pearson and Spearman tests, the human scores showed significant positive correlations. This indicates that bursts that were more single-peaked were similarly more hard-to-soft. The computational tests were less significant with respect to the Pearson tests, indicating they are not a linear correlation. However, the Spearman test was far more significant which indicates that there is a monotonic correlation between the single-peaked and hard-to-soft scores. This demonstrates that a more single-peaked burst is more hard-to-soft, albeit not linearly. 

To further analyze the data, both the human and computational scores were binned with respect to the single-peak scores. The hard-to-soft scores for each burst in a given bin were averaged and the standard error was calculated on the hard-to-soft scores. The human scores were binned into two bins corresponding to single-peaked and multi-peaked scores. Single-peaked scores were any burst with a score greater than 0.667, multi-peak bursts were the remaining bursts. The computational rankings were binned unevenly. Looking at the unnormalized single-peak scores for the horizontal line test, there was a clumping of bursts with unnormalized scores of three or less. This corresponds to final scores of at least 0.9983. Any burst with a computational single-peak score less than this was considered multi-peaked.

The number of combined uncertainties between the two bins were calculated for both the human and computational tests. The results are shown in the last column of Table \ref{tab:hts vs fred tab results}. For both the human and computational tests there is a significant difference between the hard-to-soft scores for the single-peaked and multi-peaked bursts. In particular, in both cases the single-peaked bursts had significantly larger hard-to-soft scores as compared to multi-peaked bursts. Along with the results of the Pearson and Spearman scores, this associates single-peaked behavior with hard-to-soft behavior.

%%%%%%%%%%%%%%%%%%%%%%%%%%%%%%%%%%%%%%%%%%%%%%%%%%
\section{Summary and discussion}
\label{discussion}

We have analyzed a set of Fermi GBM bright gamma-ray burst for which time-resolved spectroscopy was available \citep{Yu:2016epf}. The aim of our research was to investigate and quantify whether single-pulsed events have a stronger tendency to be characterized by the so-called hard-to-soft spectral evolution. In this case, the peak energy of the burst photons' spectra decrease monotonically in time, irrespective of the light curve behavior. Burst spectra and light curves were categorized in terms of their being single peaked and having a hard-to-soft evolution both with specifically designed software and by human interviews. While qualitative evidence of such a behavior has been cited in recent literature (e.g., \citealt{Lu2012,Basak2014,Yu:2016epf,Yu2019,Wang2024}) this is, to our best knowledge, the first attempt at a comprehensive and quantitative study. 

We find that the human and software-based classification of single-pulse and hard-to-soft behaviors are highly correlated with each other. We also find statistically significant evidence that single-peaked bursts have spectral evolution predominantly characterized as hard-to-soft. The statistical significance of this finding is stronger for the human classification, especially with the Pearson test. This result supports the model by \cite{Lazzati:2022mgg}, in which burst that explode in very dense environments --- like inside the accretion disks of supermassive black holes --- are single-pulsed and display coherent hard-to-soft evolution.

Despite the strong ($>4 \sigma$) statistical significance, our results do not support a unique identification of a spectral behavior with a light curve class. As shown in Figures~\ref{fig:fhtshumanunbinned} and~\ref{fig:fhtscompunbinned} there are single-peaked bursts with non hard-to-soft behavior as well as multi-peaked events with consistently decreasing peak energy. This should not be surprising and may be due to multiple reasons. On the one hand, there may intrinsically be bursts with these different properties. In addition, our analysis is by no means exhaustive due to data limitations. Consider for example a burst with multiple peaks with decreasing intensity (the first pulse is the brightest and the last the dimmest). This is not uncommon in multi-peaked bursts. If, however, the time resolved spectral analysis is carried out in such a way to have one spectrum for each pulse, it would be likely to see a monotonic trend in the peak frequency as well. In addition, a burst with sparse time-resolved spectroscopy may show hard-to-soft behavior accidentally. Alternatively, a bursts that displays a single peak as the result of fusing many sub-peaks together could have a non-monotonic spectral evolution but be categorized as a single-pulse event. Finally, only the best spectral fits were available meaning some bursts had different spectra functional fits for different time intervals. The peak energy values between different fits may differ slightly, contributing an additional source of noise to the results. Given the relatively small size of our sample, it is not possible to investigate further the origin of the correlation that we found, nor elaborate in detail on the source of contaminating events. Further research may include a larger burst sample and/or spectral intervals that are specifically designed to test the \cite{Lazzati:2022mgg} model. Alternatively, one may look at positional coincidence with the center of the host galaxies, like in the still unique case of GRB~191019A \citep{lazzati2023}.

\software{Python (https://www.python.org/)}

%%%%%%%%%%%%%%%%%%%%%%%%%%%%%%%%%%%%%%%%%%%%%%%%%%
\acknowledgements We would like to thank the referee for their careful and insightful comments that led to an improved manuscript. We thank Giancarlo Ghirlanda and Rosalba Perna for useful discussions. DL acknowledges support from NSF grant AST-1907955.

%

%%%%%%%%%%%%%%%%%%%%%%%%%%%%%%%%%%%%%%%%%%%%%%%%%%
\bibliographystyle{aasjournal}
\bibliography{msbiblio}

%%%%%%%%%%%%%%%%%%%%%%%%%%%%%%%%%%%%%%%%%%%%%%%%%%
\end{document}